\documentclass[letterpaper,titlepage,10pt]{article}

\usepackage{hyperref}

\usepackage{amssymb,amsmath,amsfonts}
\usepackage{epsfig}

\def\clock{{\count0=\time
           \divide\count0 60
           \ifnum\count0<10 0\fi\the\count0
           \multiply\count0 -60 \advance\count0 \time
           :\ifnum\count0<10 0\fi \the\count0
         }}
\newcommand{\timestamp}{{\small\vbox{\hbox{\tt\jobname.tex}
\hbox{\the\day/\the\month/\the\year, \clock}}}}

\newcommand{\CO}{\mathcal{O}}
\newcommand{\CN}{\mathcal{N}}

\newcommand{\Z}{\mathbb{Z}}
\newcommand{\C}{\mathbb{C}}
\newcommand{\R}{\mathbb{R}}

\newcommand{\spa}{\ , \ \ }

\newcommand{\ds}{\displaystyle}

\newcommand{\ads}{\mbox{AdS}}

\begin{document}

\begin{titlepage}

\ \
 \vskip 2 cm

\centerline{\LARGE \bf Finite size Giant Magnons in the string dual
} \vskip 0.2cm \centerline{\LARGE \bf of $\CN=6$ superconformal
Chern-Simons theory} \vskip 1.7cm

\centerline{\large {\bf Gianluca Grignani$\,^{1}$}, {\bf Troels
Harmark$\,^{2}$}, {\bf Marta Orselli$\,^{2}$} and {\bf Gordon W.
Semenoff$\,^{3}$}}

\vskip 0.5cm

\begin{center}
\sl $^1$ Dipartimento di Fisica, Universit\`a di Perugia,\\
I.N.F.N. Sezione di Perugia,\\
Via Pascoli, I-06123 Perugia, Italy\\
\vskip 0.4cm
\sl $^2$ The Niels Bohr Institute  \\
\sl  Blegdamsvej 17, 2100 Copenhagen \O , Denmark \\
\vskip 0.4 cm \sl $^3$Department of Physics and Astronomy,
University of
British Columbia\\
\sl Vancouver, British Columbia, V6T 1Z1 Canada\\
\end{center}
\vskip 0.5cm

\centerline{\small\tt grignani@pg.infn.it, harmark@nbi.dk,
orselli@nbi.dk, gordonws@phas.ubc.ca}

\vskip 1.5cm

\centerline{\bf Abstract} \vskip 0.2cm \noindent We find the exact solution for a finite size Giant
Magnon in the $SU(2)\times SU(2)$ sector of the string dual of the $\CN=6$ superconformal
Chern-Simons theory recently constructed by Aharony, Bergman, Jafferis and Maldacena. The finite size
Giant Magnon solution consists of two magnons, one in each $SU(2)$. In the infinite size limit this
solution corresponds to the Giant Magnon solution of arXiv:0806.4959. The magnon dispersion relation
exhibits finite-size exponential corrections with respect to the infinite size limit solution.


\end{titlepage}

\pagestyle{plain} \setcounter{page}{1}


\section{Introduction and summary}
\label{sec:intro}

Recently, motivated by the possible description of the worldvolume
dynamics of coincident membranes in M-theory, a new class of
conformal invariant, maximally supersymmetric field theories in 2+1
dimensions has been found \cite{Schwarz:2004yj,Aharony:2008ug}.
These theories contain gauge fields with Chern-Simons-like  kinetic
terms. Based on this development, Aharony, Bergman, Jafferis and
Maldacena proposed
 a new gauge/string duality between an $\CN=6$
 super-conformal Chern-Simons theory (ABJM theory) and type IIA string theory
 on $\ads_4 \times \C P^3$ \cite{Aharony:2008ug}. This is conjectured to constitute a new exact
 duality between gauge and string theory in addition to the celebrated duality between $\CN=4$
 superconformal Yang-Mills (SYM) theory and type IIB string theory on $\ads_5\times S^5$.

The ABJM theory consists of two Chern-Simons theories of level $k$
and $-k$ and each with gauge group $SU(N)$. It has two pairs of
chiral superfields transforming in the bifundamental representations
of $SU(N) \times SU(N)$. The R-symmetry is $SU(4)$ in accordance
with the $\CN=6$ supersymmetry of the theory. It was observed in
\cite{Aharony:2008ug} that one can define a 't Hooft coupling
$\lambda = N/k$. In the 't Hooft limit $N\rightarrow \infty$ with
$\lambda$ fixed one has a continuous coupling $\lambda$ and the ABJM
theory is weakly coupled for $\lambda \ll 1$. The ABJM theory is
conjectured to be dual to M-theory on $\ads_4\times S^7 / \Z_k$ with
$N$ units of four-form flux which for $k \ll N \ll k^5$ can be
compactified to type IIA string theory on $\ads_4 \times \C P^3$.

In the $\ads_5 / \mbox{CFT}_4$ duality major progress has been
achieved in following the tantalizing idea that the planar limit of
${\cal N}=4$ Yang-Mills theory and its string dual, the type IIB
string theory on $AdS_5\times S^5$ background, might be integrable
models which could be completely solvable using a Bethe
ansatz~\cite{Minahan:2002ve,Beisert:2003tq,Beisert:2003yb}. This
brings naturally the hope that also the new $\ads_4 / \mbox{CFT}_3$
duality can be solvable using a Bethe
ansatz~\cite{Minahan:2008hf,Gaiotto:2008cg}. However, as shown in
\cite{Grignani:2008is} this could be a more challenging task than
for $\ads_5 / \mbox{CFT}_4$ since the magnon dispersion relation in
the $SU(2)\times SU(2)$ sector of ABJM theory is shown to contain a
non-trivial function of $\lambda$, interpolating between weak and
strong coupling. A fundamental consequence of having a Bethe ansatz
is that it has distinct quasi-particles, the magnons.

In \cite{Grignani:2008is} the question of the magnon dispersion
relation was considered both from the point of view of a sigma-model
limit, a Penrose limit (see also
\cite{Nishioka:2008gz,Gaiotto:2008cg}), and furthermore using a new
Giant Magnon solution (see also \cite{Gaiotto:2008cg}). All this was
done in the $SU(2)\times SU(2)$ sector of ABJM theory, corresponding
to two two-spheres $S^2$ in the $\C P^3$ space. Adding the weak
coupling result of \cite{Minahan:2008hf,Gaiotto:2008cg} and assuming
that the symmetry arguments of \cite{Beisert:2005tm} also can be
applied to the $\ads_4 / \mbox{CFT}_3$ duality, the following
dispersion relation was found \cite{Gaiotto:2008cg,Grignani:2008is}
\begin{equation}
\label{dispgen} \Delta = \sqrt{ \frac{1}{4} + h(\lambda) \sin^2
\Big( \frac{p}{2} \Big) } \spa h(\lambda) = \left\{ \begin{array}{c}
\ds 4\lambda^2 + \CO ( \lambda^4 )
 \ \mbox{for} \ \lambda \ll 1 \\[4mm] \ds 2 \lambda + \CO ( \sqrt{\lambda} )
 \ \mbox{for} \ \lambda \gg 1 \end{array} \right.
\end{equation}

In this paper we investigate further the integrability of the
$\ads_4 / \mbox{CFT}_3$ correspondence by constructing a new finite
size Giant Magnon solution for type IIA string theory on
$\ads_4\times \C P^3$.

The Giant Magnon solution found in \cite{Grignani:2008is} is a soliton of the world-sheet sigma model
(living on $\R\times S^2\times S^2$) whose image in spacetime is a string which is pointlike in
$AdS_4$ and which rotate uniformly around the two $S^2$'s with open endpoints moving at the speed of
light on the equators of the two $S^2$'s. The solution corresponds to one of the fundamental
excitations of the spin chain with alternating sites between the fundamental and anti-fundamental
representations of the gauge theory scalar fields and next to nearest neighbor interactions, found
in~\cite{Minahan:2008hf,Gaiotto:2008cg}. The string orientation on the two $S^2$ is opposite, so that
the Giant Magnon solution can be interpreted as two giant magnons moving with equal momenta with the
same polar angle and opposite azimuthal angle.

In the infinite volume limit, integrability implies scattering with
a factorized $S$-matrix. The Bethe equation are then of the
asymptotic type, but eventually an important problem that the
integrability program would have to address is that of finite size
corrections. In this paper we address this problem and derive
exactly the conserved charges at finite size for the Giant Magnon of
the type IIA string theory on $AdS_4\times \C P^3$.

Finite size corrections to the Giant Magnon dispersion relation were
first found by generalizing Hofman and Maldacena's Giant Magnon
solution in the type-IIB sigma model on $AdS_5\times S^5$ to the
case where the size is
finite~\cite{Arutyunov:2006gs,Astolfi:2007uz}.%
\footnote{See also \cite{Minahan:2008re} for the case of an arbitrary number of Giant Magnons.} In
\cite{Astolfi:2007uz} in particular it was shown that the finite size Giant Magnon becomes a physical
string configuration, once defined on a $Z_M$ orbifold of $S^5$. The quantization of this Giant
Magnon away from the infinite size limit was discussed in \cite{Ramadanovic:2008qd} where it was
argued that this quantization inevitably leads to string theory on a $Z_M$-orbifold of $S^5$. We
shall show that also for the $AdS_4\times\C P^3$ Giant Magnon it would be possible to identify the
string endpoints by considering an orbifold of $\C P^3$~\cite{Benna:2008zy}. The orbifold
identification makes of this a legitimate closed string solution, as
in~\cite{Astolfi:2007uz,Ramadanovic:2008qd} for the $\ads_5\times S^5$ Giant Magnon.

Computing the magnon spectrum in an asymptotic expansion about
infinite size, we find that the dispersion relation, up to the
leading exponential correction, is
\begin{equation}\label{spectrum2}
\Delta-J = 2\sqrt{2\lambda}\left|\sin\frac{p}{2}\right|
-8\sqrt{2\lambda}\left|\sin\frac{p}{2}\right|^3 e^{-2-J/(\sqrt{2
\lambda}|\sin{\frac{p}{2}}|)}+\ldots
\end{equation}
where $p$ is the magnon momentum on each of the two $S^2$. The finite size corrections are
exponentially small with large $J=\frac{J_1-J_3}{2}$, where $J_1$ and $J_3$ are the generators of the
azimuthal translations on the two two-spheres.

Finite size correction to the magnon dispersion relation on $AdS_5
\times S^5$ have been reproduced from the gauge theory side using
generalized L\"uscher formulas for finite size
corrections~\cite{Janik:2007wt}. The result agrees with the
classical string computation
of~\cite{Arutyunov:2006gs,Astolfi:2007uz}. It would be extremely
interesting to make the same comparison in the case of the $\ads_4 /
\mbox{CFT}_3$ duality. However, the difference between the more
standard $\ads_5 / \mbox{CFT}_4$ duality and the $\ads_4 /
\mbox{CFT}_3$ duality is that the latter only possesses 24
supersymmetries. Therefore the checks of $\ads_4\times \C P^3$ might
indeed be more challenging than those for $\ads_5 / \mbox{CFT}_4$.

The Hofman-Maldacena Giant Magnon is a $\frac{1}{2}$-BPS state and
as such it has been shown to be part of a 16 dimensional short
multiplet of the $SU(2|2)\times SU(2|2)$
symmetry~\cite{Minahan:2007gf,Ramadanovic:2008qd}. It would be very
interesting to study and describe the supersymmetry properties of
the newly found Giant Magnon solution~\cite{Grignani:2008is} and of
its finite size version derived here. An interesting question in
fact is what happens to the supersymmetry in finite volume when a
magnon is present.  As argued in~\cite{Ramadanovic:2008qd} an
orbifold projection breaks at least half of the supersymmetry of the
$AdS_5\times S^5$ background. A string theory with a finite size
Giant Magnon therefore cannot have the same number of
supersymmetries of the parent type theory -- and could even have no
supersymmetry at all. As in~\cite{Minahan:2007gf} the study of the
existence of fermion zero modes for the fermion fluctuations of the
Green-Schwarz string sigma model on $\ads_4 /
\mbox{CFT}_3$~\cite{Arutyunov:2008if,Stefanski:2008ik} at infinite
or at finite size, might shed some light on these questions.

\section{The classical solution}

To find the Giant Magnon solution on $AdS_4\times \C P^3$ we
consider the string sigma model on this background. The coordinates
can be taken as a 5-vector $Y$ and an 8-vector $X$ where $X\in S^7$,
$Y\in AdS_4$ constrained by
\begin{eqnarray}\label{s7}&& X^2=
\sum_{i=1}^8X_iX_i=1~,~~~Y^2= \sum_{i=1}^3 Y_i^2-Y_4^2 -Y_5^2=-1\\
\label{cp3t} &&C_1=\sum_{i=1,3,5,7}\left(X_i\partial_t
X_{i+1}-X_{i+1}\partial_t X_i\right)=0~,~~~
C_2=\sum_{i=1,3,5,7}\left(X_i\partial_s X_{i+1}-X_{i+1}\partial_s
X_i\right)=0\cr&&
\end{eqnarray}
The constraints $C_1=0$ and $C_2=0$ define the background to be $\C
P^3$.

 The bosonic part of the sigma model action in the conformal gauge is
\begin{equation}
S= -\sqrt{2\lambda}\int dt \int ds \left[ \frac{1}{4}\partial_a
Y\cdot \partial^a Y +
\partial_a X\cdot \partial^a X + \tilde \Lambda(Y^2+1)+\Lambda(X^2-1)+\Lambda_1 C^2_1+\Lambda_2 C^2_2\right]
\label{action}\end{equation}

The coupling constant of the sigma model is the inverse radius of
curvature squared of the constant curvature spaces
$\frac{1}{R^2}=\frac{1}{4\pi\sqrt{2\lambda}}$. The worldsheet metric
has signature $(-,+)$ such that
$\partial_a\partial^a=-\frac{\partial^2}{\partial
t^2}+\frac{\partial^2}{\partial s^2} =-\partial_+\partial_-$ and
$\sigma^{\pm}=\frac{1}{2}\left(t\pm s\right)$. Here, $\Lambda$,
$\tilde\Lambda$ and $\Lambda_i$, $i=1,2$ are Lagrange multipliers
which enforce the coordinate constraints (\ref{s7}) and
(\ref{cp3t}). In closed string theory, the worldsheet has
cylindrical topology. The range of the time coordinate $t$ is
infinite and the range of the space coordinate is taken to be $s\in
(-r,r]$.  The parameter $r$ can be changed by scaling the worldsheet
coordinates. We shall later fix $r$ to a convenient value. The
equations of motion following from the action (\ref{action}) should
be supplemented by Virasoro constraints,
\begin{eqnarray}\label{vir1}
\partial_+X\cdot\partial_+X+\frac{1}{4}\partial_+Y\cdot\partial_+Y=0
~~,~~
\partial_-X\cdot\partial_-X+\frac{1}{4}\partial_-Y\cdot\partial_-Y=0
\end{eqnarray}
The Giant Magnon solution will be found as a solution of the
classical equations of motion where only coordinates on two
$S^2\subset S^7$ and $R^1\subset AdS_7$ are excited.  The solution
on $AdS_5\times S^5$ was originally found by Hofman and
Maldacena~\cite{Hofman:2006xt} in the limit where $r$ is infinite.
This is a closed string solution with open boundary conditions in
one azimuthal direction.

In the case we are studying the solution is point-like in $AdS_4$
and extended along the two $S^2$ which are subsets of $S^7$. The
solution lives on an $R^1\times S^2\times S^2$ subspace of
$AdS_4\times S^7$, the $R^1\subset AdS_7$ and $S^2\times S^2\subset
S^7$. We shall choose the solution in such a way that it has
opposite azimuthal angles on the two $S^2$ and the same polar
angles. The boundary conditions are those of closed string theory.
All variables are periodic, except for the azimuthal angles of the
two $S^2$'s which will be chosen to obey the magnon boundary
condition which on one $S^2$ is
\begin{equation}\label{magnonboundarycondition1}
\Delta \phi_1\equiv p\end{equation} and on the other one will be
\begin{equation}\label{magnonboundarycondition2}
\Delta \phi_2 \equiv-p\end{equation} These identifications corresponds to opposite orientations of
the string on the two $S^2$. The Giant Magnon is then characterized by the momentum $p$ and by the
choice of the point in the transverse directions to the two $S^2$, $i.e.$ by 2 two-component
polarization vectors. $p$ has to be interpreted as the momentum of the magnons in the spin chain,
these two magnons have equal magnon momentum. They give the same contribution to the total momentum
constraint.

We begin with the ansatz that the solution lives on an $R^1\times
S^2\times S^2$ subspace of $AdS_4\times S^7$, with
$\phi_1=-\phi_2=\phi$ and $\theta_1=\theta_2=\theta$, thus the
ansatz is
\begin{eqnarray}\label{magnonansatz0}
&&Y_4+iY_5 = e^{i \frac{\Delta}{r\sqrt{2\lambda}} t}~~,~~~~Y_1=Y_2=Y_3=0 \\
\label{magnonansatz} &&X_1+iX_2 = \frac{1}{\sqrt{2}}e^{
i\phi(t,s)}\sqrt{1-z^2(t,s)}~,~~~~X_5+iX_6 = \frac{1}{\sqrt{2}}e^{-
i\phi(t,s)}\sqrt{1-z^2(t,s)}~~\cr &&(X_3,X_4)=\frac{\hat
n_1}{\sqrt{2}} z(t,s)~~,~~~~(X_7,X_8)=\frac{{\hat n_2}}{\sqrt{2}}
~z(t,s)
\end{eqnarray}
where $\hat n_i$ $i=1,2$ are constant unit vectors and $z(t,s)=\cos\theta(t,s)$   is a function
taking values on the interval $0\leq z < 1$. The ansatz (\ref{magnonansatz0})-(\ref{magnonansatz}) is
such that the constraints (\ref{s7}) and (\ref{cp3t}) are automatically satisfied. The boundary
condition for the magnon is
\begin{equation}\label{wrappedstringboundarycondition}
\phi(t,s=r)-\phi(t,s=-r)~=~p~~~
\end{equation}
with all other variables periodic.

Using \eqref{wrappedstringboundarycondition} we see that the boundary conditions of the $X_i$
variables at finite size are
\begin{eqnarray}\label{bcsfs} &&X_1 + i X_2 \large|_{s=r}=e
^{i p}\left(X_1 + i X_2\right)\large|_{s=-r}~,~~~~X_5 + i X_6 \large|_{s=r}=e ^{-i p}\left(X_5 + i
X_6\right)\large|_{s=-r}\cr &&(X_3,X_4)\large|_{s=r}=(X_3,X_4)\large|_{s=-r}~,~~~~
(X_7,X_8)\large|_{s=r}=(X_7,X_8)\large|_{s=-r}
\end{eqnarray}
For $s\to\infty$ we get from \eqref{bcsfs} the boundary conditions of the solution found
in~\cite{Grignani:2008is}. The $\C P^3$ corresponds to making the identification
\begin{equation}
X_{2j-1} + i X_{2j} = Z ( \hat{X}_{2j-1} + i \hat{X}_{2j} )\spa j=1,\dots, 4
\end{equation}
where $Z \in \C$ and $X_i$ and $\hat{X}_i$ identify two points on $\C^4$. We see that the two string
endpoints at $s = \pm r$ are {\sl not} identified under this. Quite remarkably, however, it is
possible to find a setting where the boundary conditions \eqref{bcsfs} correspond to identified
endpoints. In ref.~\cite{Benna:2008zy} orbifold projections of the ABJM theory were considered. These
give non-chiral and chiral $(U(N)\times(N))^n$ superconformal quiver gauge theories. These theories
at level $k$ are dual to certain $AdS_4 \times S^7/(Z_M \times Z_k)$ backgrounds of $M$-theory. In
particular in ref.~\cite{Benna:2008zy} an orbifold projection of the non-chiral ABJM theory, which
produces a chiral gauge theory, was considered. This was done by placing $N$ $M_2$-branes at the
singularity of $\C^4/(Z_M\times Z_k)$, where the $Z_M$ action is given by
\begin{eqnarray}\label{bcsk}
&&X_1 + i X_2 \to e ^{\frac{2\pi i m}{M}}\left(X_1 + i X_2\right)~,~~~~X_5 + i X_6 \to e
^{-\frac{2\pi i m}{M}}\left(X_5 + i X_6\right)\cr &&X_3+i X_4\to X_3+i X_4~,~~~~ X_7+iX_8\to X_7+iX_8
\end{eqnarray}
These identifications are identical to those produced by the Giant Magnon boundary conditions
\eqref{bcsfs} if we chose $p=\frac{2\pi m}{M}$. We can thus conclude that, as for the $AdS_5\times
S^5$ finite size giant magnon~\cite{Arutyunov:2006gs,Astolfi:2007uz}, considering an orbifold of the
original theory~\cite{Astolfi:2007uz}, the endpoints of the string are identified, $i.e.$ the
orbifold group acts in such a way  that it identifies the ends of the string, resulting in a
legitimate state of closed string theory. This was advocated in~\cite{Astolfi:2007uz} as a way to
study the spectrum of a single magnon in a setting, $AdS_5\times S^5/Z_M$, where it is a physical
state and there are no issues of gauge invariance~\cite{Arutyunov:2006gs}. The same thing seems to
happen also for the $AdS_4\times \C P^3$ magnon, a natural setting for giving physical sense to this
solution as a closed string state is to put it on an orbifold. We could then argue as
in~\cite{Ramadanovic:2008qd} that if we consider the giant magnon at finite size as a quantum string
state, with the boundary condition that the string is open in the direction of the magnon motion, we
are $inevitably$ led to an orbifold.

The solution on $AdS_4$ in (\ref{magnonansatz0}) is chosen so that the energy density is constant and
the total energy of the string is the integral
\begin{equation}
\Delta = -\frac{\sqrt{2\lambda}}{2}\int_{-r}^r ds\left[ Y_4\dot
Y_5-Y_5\dot Y_4\right]
\end{equation}  With the ansatz (\ref{magnonansatz0}) and (\ref{magnonansatz}) the action reduces
to
\begin{equation}
S=\sqrt{2\lambda}\int dt \int_{-r}^r ds \left[ (1-z^2)\partial_+
\phi\partial_-\phi +\frac{
\partial_+ z\partial_- z}{1-z^2}\right]
\end{equation}
 up to a constant.
The equations of motion are
\begin{eqnarray}\label{eqmophi}
&&\partial_+\left( (1-z^2)\partial_-\phi\right)+
\partial_-\left( (1-z^2)\partial_+\phi\right)=0
\\
\label{eqmoz} &&\partial_+\left( \frac{\partial_-z}{1-z^2}\right)+
\partial_-\left( \frac{\partial_+z}{1-z^2}\right)
=\frac{
2z\partial_+z\partial_-z}{(1-z^2)^2}-2z\partial_+\phi\partial_-\phi
\end{eqnarray}
and the Virasoro constraints are
\begin{eqnarray}\label{v1}
&& T_{++}=(1-z^2) \partial_+\phi\partial_+\phi +
\frac{\partial_+z\partial_+z }{1-z^2}-
\frac{1}{4}\left(\frac{\Delta}{r\sqrt{2\lambda}}\right)^2\sim
0\\&&\label{v2}T_{--}= (1-z^2)
\partial_-\phi\partial_-\phi + \frac{\partial_-z\partial_-z }{1-z^2}-
\frac{1}{4}\left(\frac{\Delta}{r\sqrt{2\lambda}}\right)^2\sim 0
\end{eqnarray}

We will choose the parameter $r$ so that
\begin{equation}\label{r}
r=\frac{\Delta}{2\sqrt{2\lambda}}
\end{equation} this simplifies the constraints
(\ref{v1}) and (\ref{v2}) so that the last term in each expression
is equal to 1. In the standard solution $r$ would not contain the
factor of 1/2. It is easy to check that the Virasoro constraints are
compatible with the equations of motion
\begin{equation}\label{conservationofemtensor}
\partial_-T_{++}=0 ~~,~~ \partial_+ T_{--}=0
\end{equation}

Now, we make the ansatz that the Giant Magnon on $S^2$ is a
right-moving soliton
\begin{equation}\label{periodicansatz}
\phi(t,s)= \Psi t+\Omega s+\varphi(u)
~~,~~z(t,s)=z(u)\end{equation} where we use the boosted variables
\begin{equation}\label{boostedvariables} \left[\begin{matrix} u \cr v \cr
\end{matrix} \right] = \left[ \begin{matrix} \cosh\eta & -\sinh \eta \cr -\sinh\eta
&
    \cosh\eta \cr \end{matrix} \right] \left[\begin{matrix} s \cr t \cr \end{matrix}\right]
    \end{equation} In (\ref{periodicansatz}) we have allowed for
 time-dependence of the angle $\phi(t,s)$ with $\Psi t$, taken into account the boundary condition
 (\ref{wrappedstringboundarycondition}) with $\Omega s$ where
\begin{equation}\label{definitionofOmega}\Omega~=~
\frac{p}{2r}
\end{equation} and we now assume that the remaining functions $\varphi(u)$
and $z(u)$ are periodic,
\begin{equation}\label{periodicboundaryconditions}\varphi(u+2r\cosh\eta)=
\varphi(u)~~,~~z(u+2r\cosh\eta)=z(u)
\end{equation}
This implies the identities $\int_{-r}^r ds
\dot\varphi=0=\int_{-r}^r ds \varphi'$ which we shall use later.
From now on, over-dot will denote $\frac{d}{du}$.

With the ansatz (\ref{periodicansatz}) the equations of motion
(\ref{eqmophi}) for $\phi$ becomes
\begin{equation}
\frac{d}{du}\left( (1-z^2)\left(\dot\varphi
+\Psi\sinh\eta+\Omega\cosh\eta\right)\right)=0
\end{equation}
This equation implies that the quantity in front of the derivative
is a constant, which we shall denote as $j$. Then
\begin{eqnarray}\label{theconstantj}
\dot\varphi = \frac{j}{1-z^2}-\Psi\sinh\eta -\Omega\cosh\eta
\end{eqnarray}

With the anzatz (\ref{periodicansatz}), the equations of motion
(\ref{eqmophi}) and (\ref{eqmoz}) are second order differential
equations  for the functions $z(u)$ and $\varphi(u)$ of the variable
$u$. Since they now have one variable, they are equivalent to the
conservation of the energy-momentum tensor, i.e.~the equations
(\ref{conservationofemtensor}). For this reason, the Virasoro
constraints (\ref{v1}) and (\ref{v2}) with the ansatz substituted
are a first integral of the equations of motion. With the magnon
ansatz (\ref{periodicansatz}), they are
\begin{eqnarray}
(1-z^2)\left( \Psi-\Omega-e^{\eta}\dot\varphi\right)^2 +
e^{2\eta}\frac{ \dot
z^2}{1-z^2}=1 \\
(1-z^2)\left( \Psi + \Omega+e^{-\eta} \dot \varphi\right)^2 +
e^{-2\eta}\frac{ \dot z^2}{1-z^2}=1
\end{eqnarray}
These equations are compatible with (\ref{theconstantj}) and give
the equations for the parameters which determines $j$:
\begin{equation}
j= \frac{\sinh2\eta}{2 \left(\Psi\cosh\eta +\Omega \sinh \eta\right)
}
\end{equation}
and the equations which determines $\dot z(u)$:
\begin{equation}
\left( \frac{dz}{du}\right)^2 =  \frac{\left(z^2-z^2_{\rm
   min}\right)\left( z^2_{\rm max}-z^2\right) }{z^2_{\rm
 max}-z^2_{\rm min} }
\label{eqmo}\end{equation} where the turning points are
\begin{equation}
z^2_{\rm max}=1-\frac{ \sinh^2\eta}{\left(\Psi\cosh\eta +
\Omega\sinh\eta\right)^2}
\end{equation}
\begin{equation}
z^2_{\rm min}=1-\frac{\cosh^2\eta}{\left(\Psi\cosh\eta +
\Omega\sinh\eta\right)^2}
\end{equation}
These imply
\begin{eqnarray} \label{hyperbolicfunctions}
&&\cosh\eta = \sqrt{ \frac{ 1-z^2_{\rm min} }{z^2_{\rm max} -
z^2_{\rm
   min}}}
~~,~~ \sinh\eta = \sqrt{ \frac{ 1-z^2_{\rm max} }{z^2_{\rm max} -
z^2_{\rm
   min}}} \\ \label{upsilonandomega}
&&\Psi\cosh\eta+\Omega\sinh\eta = \frac{1}{\sqrt{z_{\rm
max}^2-z_{\rm
   min}^2}}
\end{eqnarray}
The solution is obtained by integrating (\ref{eqmo}),
\begin{eqnarray}\label{u}
u=-\int_{z_{\rm max}}^{z(u)} dz \frac{ \sqrt{
 z^2_{\rm max} - z^2_{\rm min} }}{ \sqrt{ z^2 - z^2_{\rm
   min}}   \sqrt{ z^2_{\rm max} - z^2} }~~,~~ u>0 \\
u= -\int^{z_{\rm max}}_{z(u)} dz \frac{ \sqrt{
 z^2_{\rm max} - z^2_{\rm min} }}{ \sqrt{ z^2 - z^2_{\rm
   min}}   \sqrt{ z^2_{\rm max} - z^2} }~~,~~ u<0
\end{eqnarray}
We have chosen the constant of integration so that the maximum of
$z(u)$, $z_{\rm max}$ occurs at $u=0$ and the minimum is at $u=\pm
r\cosh\eta$.  $\frac{dz}{du}$ is positive when $u<0$ and negative
when $u>0$. The resulting solutions are even functions of $u$,
$z(u)=z(-u)$. The result of the integrals in (\ref{u}) are the
incomplete elliptic integrals of the first kind.
\begin{eqnarray}
u=
{\nu}\int_0^{\hat\theta(z)}\frac{d\theta}{\sqrt{1-\nu^2\sin^2\theta}}
={\nu}F\left( \hat\theta(z) ,\nu\right) ~~,~~ 0\leq u\leq r\cosh\eta\\
u=-{\nu}\int_0^{\hat\theta(z)}\frac{d\theta}{\sqrt{1-\nu^2\sin^2\theta}}
=- {\nu}F\left( \hat\theta(z) ,\nu\right) ~~,~~ -r\cosh\eta\leq u
\leq 0
\end{eqnarray}
where
\begin{equation}
\hat\theta(z) =\arcsin\sqrt{\frac{z_{\rm maz}^2-z^2}{z_{\rm
     max}^2-z_{\rm min}^2}} ~~,~~ \nu=\sqrt{1-\frac{z^2_{\rm min}
}{z^2_{\rm max}}}
\end{equation}
and we are using the standard notation for the arguments of the
elliptic function given in Ref.~\cite{Grad}. The argument of the
function is $z(u)$ which is then given by a Jacobi elliptic
function,
\begin{equation}\label{inverselliptic}
z(u)= z_{\rm max}{\rm \bf dn} \left( \frac{u }{ {\nu}},\nu\right)
\end{equation}
 It is the finite size Giant Magnon
solution, given in terms of two integration constants $z_{\rm max}$
and $z_{\rm min}$. In the next subsection we will discuss how these
constants can be determined in terms of the energy and angular
momentum of the solution.

\subsection{Constants of integration}

We note that the length of the worldsheet is
\begin{eqnarray}
&& r=\int_0^r ds =-\frac{1}{\cosh\eta}\int_{z_{\rm max}}^{\rm z_{\rm
min}}dz \frac{du}{dz}\cr&& = \frac{1}{\cosh\eta} \int_{z_{\rm
min}}^{z_{\rm max}}
 dz \frac{\sqrt{z_{\rm max}^2-z_{\rm min}^2}}{\sqrt{z_{\rm
       max}^2-z^2}\sqrt{z^2-z^2_{\rm min}}}=\frac{ {\nu}}{\cosh\eta}K(\nu)
\end{eqnarray}
where $K(\nu)=F(\frac{\pi}{2},\nu)$ is the complete Elliptic
integral of the first kind (see Appendix.\ref{app:Elliptic}).
Remembering that $r=\frac{\Delta}{2\sqrt{2\lambda}}$, eq.(\ref{r}),
we see that this yields
\begin{equation}
\Delta=2\sqrt{2\lambda} \left(\frac{z_{\rm max}^2-z_{\rm min}^2}
{z_{\rm max}\sqrt{1-z_{\rm
min}^2}} K(\nu)\right)~~,~~\nu=\sqrt{1-\frac{z_{\rm min}^2}{z_{\rm max}^2}} \label{equationfordelta}\\
\end{equation}
  Also by relating the length of the worldsheet to $(z_{\rm max},z_{\rm
min})$, it ensures that the period of the inverse elliptic function
in (\ref{inverselliptic}) is the correct one.

Next, we shall derive the equation for the world sheet momentum. In
Eq.~(\ref{periodicansatz}), the zero-modes proportional to $\Psi$
and $\Omega$ were separated so that the remaining function
$\varphi(u)$ is periodic in $u$.  This implies that
$\int_{-r\cosh\eta}^{r\cosh\eta} du \dot\varphi(u)=0$.
Eq.~(\ref{theconstantj}) determines its derivative
$\frac{d\varphi}{du}$ in terms of $z(u)$ and constants as $
\frac{d}{du}\varphi(u) = \frac{j}{1-z^2} - \left(\Psi\sinh\eta +
\Omega\cosh\eta\right) $. The right-hand-side of this equation is a
periodic and even function of $u$. Then, integrating both sides over
the range of the $u$'s and using the above observation that the
integral of the left-hand-side must vanish, we find the identity
$$
j\int_0^{r\cosh\eta} du \frac{1}{1-z^2} = r\cosh\eta
\left(\Psi\sinh\eta + \Omega\cosh\eta \right)
$$
Using (\ref{theconstantj}) and (\ref{equationfordelta}), and
recalling (\ref{upsilonandomega}) we find
\begin{eqnarray}
\Psi\sinh\eta + \Omega\cosh\eta &=& \frac{1}{r}\frac{ \sqrt{z_{\rm
max}^2 - z_{\rm min}^2} } { z_{\rm max}\sqrt{1-z^2_{\rm max}} }
\Pi\left(\frac{z^2_{\rm max}-z^2_{\rm min}}{z^2_{\rm max}-1}; \nu
\right)
\nonumber \\
\Psi\cosh\eta + \Omega\sinh\eta &=& \frac{1}{ \sqrt {z_{\rm
max}^2-z_{\rm min}^2} }
\end{eqnarray}
which we can solve to get
\begin{equation}\label{intermediateequationforomega}
\Omega= \frac{1}{r} \frac{ \sqrt{ 1 - z_{\rm min}^2} } { z_{\rm max}
\sqrt{1-z^2_{\rm max} } } \Pi\left(\frac{z^2_{\rm max}-z^2_{\rm
min}}{z^2_{\rm max}-1}; \nu \right) -\frac{\sqrt{1-z_{\rm max}^2}}{
z_{\rm max}^2-z_{\rm min}^2 }
\end{equation}
where $\Pi$ is the complete elliptic integral of the third kind (see
Appendix.\ref{app:Elliptic}).

Combining Eqs.~(\ref{definitionofOmega}), (\ref{equationfordelta})
and (\ref{intermediateequationforomega}) we find
\begin{equation}
\label{equationforworldsheetmomentum} \frac{ p}{2} =\frac{\sqrt{1 -
z_{\rm min}^2}}{z_{\rm max}\sqrt{1-z_{\rm max}^2} }
~\left(\Pi\left(\frac{z^2_{\rm max}-z^2_{\rm min}}{z^2_{\rm max}-1};
\nu \right) - \frac{1-z_{\rm max}^2}{1-z_{\rm min}^2}K( \nu )
\right)
\end{equation}
In the orbifold case discussed above, see eq.\eqref{bcsk}, we should just set $\frac{p}{2}=\frac{\pi
m}{M}$, for an $m$-times wrapped string.

Finally we shall compute the angular momentum, which is given by the
Noether charge
\begin{eqnarray}
J\equiv\frac{J_1-J_3}{2} &=& -2\sqrt{2\lambda}\int_{-r}^r
ds\left[\frac{X_1\dot X_2-X_2\dot X_1}{2}-\frac{X_5\dot X_6-X_6\dot
X_5}{2}\right]=\sqrt{2\lambda}\int_{-r}^r
ds\left(1-z^2\right)\frac{d}{dt}\phi \cr
&=&2\sqrt{2\lambda}\frac{\sqrt{ z^2_{\rm max}-z^2_{\rm
min}}}{\cosh\eta}\int_{z_{\rm min}}^{z_{\rm max}} \frac{dz
(1-z^2)}{\sqrt{z^2_{\rm max}-z^2}\sqrt{z^2-z_{\rm min}^2}}\left(
\Psi -\sinh\eta\dot\varphi\right)\end{eqnarray} Then, we use
(\ref{theconstantj}), (\ref{upsilonandomega}) and (\ref{ellipticfs})
to find the identity
\begin{equation}
\label{equationforJ} J= 2\sqrt{2\lambda}z_{\rm max}\left( K(\nu) -
E(\nu)\right)
\end{equation}
where $K$, $E$ and $\Pi$ are the complete elliptic integrals of the
first, second and third kinds, respectively (see
Appendix.\ref{app:Elliptic}). Equations (\ref{equationfordelta}),
(\ref{equationforJ}) and (\ref{equationforworldsheetmomentum}) are
identical, a part for the overall factors, to those quoted in
Eqs.~(36), (37) and (38) of Ref.~\cite{Astolfi:2007uz} and, with
minor misprints corrected and $a=0$, (B.4), (B.5) and (B.6) of
Ref.~\cite{Arutyunov:2006gs}. In those works, they were found using
a light-cone gauge, and in the latter the conformal gauge and the
results for physical quantities agree with each other.

In principle, two of the equations (\ref{equationfordelta}),
(\ref{equationforJ}) and (\ref{equationforworldsheetmomentum}) can
be used to determine $z_{\rm min}$ and $z_{\rm max}$ in terms of the
target space quantities. The third then gives an equation for the
spectrum of the magnon, relating $\Delta$, $J$ and $p$. In practice,
this can be done in the limit where $\Delta$ and $J$ are large. This
limit will be discussed in the next section.

\section{The magnon limit}

The magnon limit takes $\Delta$ and $J$ large, so that $\Delta-J$
remains finite.  This is achieved by taking $z_{\rm min}\to 0$.
Using Eqs.~(\ref{asymptoticK}),(\ref{asymptoticE}) and
(\ref{asymptoticPi}), we can find an asyptotic expansion of
Eqs.~(\ref{equationfordelta}), (\ref{equationforJ}) and
(\ref{equationforworldsheetmomentum}),
\begin{eqnarray}\label{asymptoticdelta} \Delta &=&
2\sqrt{2\lambda}z_{\rm max}\left\{ \ln \frac{4z_{\rm max}}{z_{\rm
min}} + \frac{1}{4}\frac{z_{\rm min}^2}{z_{\rm max}^2}\left[
(2z_{\rm max}^2-3)\ln\frac{4z_{\rm max}}{z_{\rm min}} -
1\right]+\ldots\right\}
\\ \label{asymptoticJ} J&=& 2\sqrt{2\lambda}z_{\rm max}\left\{-1+\left(1-\frac{1}{4}\frac{z_{\rm
min}^2}{z_{\rm max}^2}\right)\ln \frac{4z_{\rm max}}{z_{\rm
min}}+\ldots\right\}
\\
\label{asymptoticworldsheetmomentum} \frac{p}{2}&=& \arcsin z_{\rm
max} -\frac{1}{4}\frac{z_{\rm min}^2}{z_{\rm max}^2} z_{\rm max}
\sqrt{ 1-z_{\rm max}^2} \left( 2 \ln\frac{4z_{\rm max}}{z_{\rm
min}}+1\right)+\ldots\end{eqnarray}

Then, in the leading order,
\begin{eqnarray}\label{solutionforzmax}
z_{\rm max}&=&\left|\sin\frac{ p}{2}\right|+\ldots
\\ \label{solutionforzmin}
z_{\rm min}&=&4\left|\sin\frac{ p}{2}\right|~\exp\left(\frac{-
\Delta}{2\sqrt{2\lambda}\left|\sin\frac{p }{2}\right|}\right)+\ldots
\approx 0
\end{eqnarray}
We have chosen the solution where $z(u)$ is a positive function and
therefore $z_{\rm max}$ and $z_{\rm min}$ are positive numbers.  The
Giant Magnon achieves maximum height $z_{\rm max}$ which is itself
maximal when $p=\pi$.  The smallest value of $z(u)$, $z_{\rm min}$,
is always smaller by a factor that is exponentially small in the
size $\Delta$ and is zero in the Giant Magnon limit.

Taking the infinite $J$ limit, we get the equation for the spectrum
obtained in \cite{Grignani:2008is}
\begin{eqnarray}
\Delta-J =  2\sqrt{2\lambda}\left|\sin\frac{p }{2}\right|+\ldots
\label{leadingordersolution}
\end{eqnarray}
From this equation we see that, for a very small magnon, $p\ll 1$,
$\Delta-J\sim\sqrt{2\lambda} p$.

We see furthermore that, to the leading order
(\ref{leadingordersolution}), it is easy to find the explicit
solution, \begin{equation} z(u)= \frac{\sin\frac{p }{2}}{\cosh u}
~~,~~ \phi(t,s)= t+ \arctan\left(\tan\frac{p}{2}\tanh
u\right)\label{magnoninthemagnonlimit}\end{equation} which using the
ansatz  (\ref{magnonansatz0})-(\ref{magnonansatz}) gives back the
infinite $J$ limit solution found in \cite{Grignani:2008is}.

Finally, the leading exponential corrections to the magnon limit are
easy to find. To the next-to-leading order we compute
\begin{equation}\label{disprelfin}
\Delta-J=2\sqrt{2\lambda}\left\{\left|\sin\frac{ p}{2}\right|-4
\left|\sin^3\frac{ p}{2}\right|
\exp\left(\frac{-\Delta}{\sqrt{2\lambda}\left|\sin\frac{p
}{2}\right|}\right)+ \ldots\right\}
\end{equation}
The exponential correction is the leading finite-size correction to the Giant Magnon dispersion
relation. For the orbifold \eqref{bcsk}, $p$ in \eqref{disprelfin} should should just be set to
$p=\frac{2 \pi m}{M}$, for an $m$-times wrapped string.

\begin{appendix}
\section{Complete Elliptic Integrals} \label{app:Elliptic}

Above we used the integral formulae for complete elliptic integrals
of the first, second and third kinds, respectively
\begin{eqnarray} \label{ellipticfs}
\int_{z_{\rm min}}^{z_{\rm max}}dz\frac{1}{\sqrt{z^2-z^2_{\rm
min}}\sqrt{z^2_{\rm max}-z^2}}&=&\frac{1}{z_{\rm max}}K(\nu)\\
\int_{z_{\rm min}}^{z_{\rm max}}dz\frac{z^2}{\sqrt{z^2-z^2_{\rm
min}}\sqrt{z^2_{\rm max}-z^2}}&=&z_{\rm max}E(\nu)\\ \int_{z_{\rm
min}}^{z_{\rm max}}\frac{dz}{(1-z^2)\sqrt{z^2-z^2_{\rm
min}}\sqrt{z^2_{\rm max}-z^2}}&=&  \frac{\Pi\left(\frac{z^2_{\rm
max}-z^2_{\rm min}}{z^2_{\rm max}-1}; \nu\right) }{z_{\rm
max}(1-z^2_{\rm max})}
\end{eqnarray}
where $\nu=\sqrt{1-\frac{z^2_{\rm min}}{z^2_{\rm max}}}$. We have
taken conventions for the arguments of these functions which are
defined by Ref.~\cite{Grad}.  In the paper we used asymptotic
expansions around the limit $z_{\rm min}\to 0$,
\begin{eqnarray}\label{asymptoticK} K(\nu)&=& \ln\left(
4\frac{z_{\rm max}}{z_{\rm min}}\right) +\frac{1}{4}\frac{z_{\rm
min}^2}{z_{\rm max}^2} \left( \ln\left( 4\frac{z_{\rm max}}{z_{\rm
min}}\right)-1\right) +\ldots \\
\label{asymptoticE} E(\nu)&=&1+ \frac{1}{4}\frac{z_{\rm
min}^2}{z_{\rm max}^2}\left(
2\ln\left( 4\frac{z_{\rm max}}{z_{\rm min}}\right)-1\right)+\ldots\\
\label{asymptoticPi}  \Pi\left(\frac{z^2_{\rm max}-z^2_{\rm
min}}{z^2_{\rm max}-1}; \nu \right) &=& \left( 1-z_{\rm
max}^2\right) \left[\ln\left( 4\frac{z_{\rm max}}{z_{\rm
min}}\right) +\frac{z_{\rm min}^2}{4z_{\rm
max}^2}\left(\left(2z_{\rm max}^2+1\right)\ln \frac{4z_{\rm
max}}{z_{\rm min}} - \left(z_{\rm max}^2+1\right)\right) \right] +
\nonumber \\ &+&\left( 1+\frac{z_{\rm min}^2}{2}\right)z_{\rm
max}\sqrt{1-z_{\rm max}^2}~\arcsin z_{\rm max}+\ldots
\end{eqnarray}
where the three dots indicate terms of order $z_{\rm min}^4\ln
z_{\rm min}$ in all cases.
\end{appendix}


\providecommand{\href}[2]{#2}\begingroup\raggedright\endgroup

\end{document}